\documentclass[%
 reprint,
superscriptaddress,
showpacs,preprintnumbers,
 amsmath,amssymb,
 aps,
prb,
floatfix,
]{revtex4-1}

\usepackage{graphicx}
\usepackage{dcolumn}
\usepackage{bm}

\usepackage{hyperref}
\hypersetup{
    unicode=false,          
    pdftoolbar=true,        
    pdfmenubar=true,        
    pdffitwindow=false,     
    pdfstartview={FitH},    
    pdftitle={My title},    
    pdfauthor={Author},     
    pdfsubject={Subject},   
    pdfcreator={Creator},   
    pdfproducer={Producer}, 
    pdfkeywords={keyword1} {key2} {key3}, 
    pdfnewwindow=true,      
    colorlinks=true,       
    linkcolor=blue,          
    citecolor=blue,        
    filecolor=blue,      
    urlcolor=blue,       
    plainpages=false        
}

\begin{document}
\title{Off-axis parabolic mirror optics for polarized Raman spectroscopy at low temperature}
\date{\today}
\author{N. Chelwani}
\affiliation{Walther Meissner Institut, Bayerische Akademie der Wissenschaften, 85748 Garching, Germany}
\affiliation{Fakult\"at f\"ur Physik E23, Technische Universit\"at M\"unchen, 85748 Garching, Germany}
\author{D. Hoch}
\affiliation{Walther Meissner Institut, Bayerische Akademie der Wissenschaften, 85748 Garching, Germany}
\affiliation{Fakult\"at f\"ur Physik E23, Technische Universit\"at M\"unchen, 85748 Garching, Germany}
\author{D. Jost}
\affiliation{Walther Meissner Institut, Bayerische Akademie der Wissenschaften, 85748 Garching, Germany}
\affiliation{Fakult\"at f\"ur Physik E23, Technische Universit\"at M\"unchen, 85748 Garching, Germany}
\author{B.~Botka}
\altaffiliation{Present address: University of South Australia, Adelaide SA 5000, Australia}
\affiliation{Walther Meissner Institut, Bayerische Akademie der Wissenschaften, 85748 Garching, Germany}
\affiliation{Wigner Research Centre for Physics, Hungarian Academy of Sciences, 1525 Budapest, Hungary}
\author{J.-R.~Scholz}
\affiliation{Walther Meissner Institut, Bayerische Akademie der Wissenschaften, 85748 Garching, Germany}
\affiliation{Fakult\"at f\"ur Physik E23, Technische Universit\"at M\"unchen, 85748 Garching, Germany}
\author{R. Richter}
\affiliation{Walther Meissner Institut, Bayerische Akademie der Wissenschaften, 85748 Garching, Germany}
\affiliation{Fakult\"at f\"ur Physik E23, Technische Universit\"at M\"unchen, 85748 Garching, Germany}
\author{M. Theodoridou}
\affiliation{Walther Meissner Institut, Bayerische Akademie der Wissenschaften, 85748 Garching, Germany}
\affiliation{Fakult\"at f\"ur Physik E23, Technische Universit\"at M\"unchen, 85748 Garching, Germany}
\author{F.~Kretzschmar}
\altaffiliation{Present address: Intel Mobile Communications, Am Campeon 10-12, 85579 Neubiberg, Germany}
\affiliation{Walther Meissner Institut, Bayerische Akademie der Wissenschaften, 85748 Garching, Germany}
\affiliation{Fakult\"at f\"ur Physik E23, Technische Universit\"at M\"unchen, 85748 Garching, Germany}
\author{T. B\"ohm}
\affiliation{Walther Meissner Institut, Bayerische Akademie der Wissenschaften, 85748 Garching, Germany}
\affiliation{Fakult\"at f\"ur Physik E23, Technische Universit\"at M\"unchen, 85748 Garching, Germany}
\author{K. Kamar\'as}
\affiliation{Wigner Research Centre for Physics, Hungarian Academy of Sciences, 1525 Budapest, Hungary}
\author{R. Hackl}
\affiliation{Walther Meissner Institut, Bayerische Akademie der Wissenschaften, 85748 Garching, Germany}

\begin{abstract}
  We report the development of a detection optics for the integration of Raman scattering and scanning probe microscopy at low temperature based on a parabolic mirror. In our set-up half of the paraboloid mirror covers a solid angle of $\pi$ corresponding to a numerical aperture of N.A.\,$\approx 0.85$. The optical system can be used for far- and near-field spectroscopy. In the far field the polarizations can be maintained to within 80-90\%. In combination with a scanning microscope (AFM/STM), tunneling or near-field experiments are possible with less than 10\% loss of aperture. Our set-up provides ideal conditions for the future development of tip-enhanced Raman spectroscopy (TERS) at low temperature.
\end{abstract}
\maketitle


\section{Introduction}
High spatial resolution and the \textit{in situ} combination of different types of spectroscopies at the exactly identical position of a sample is a key requirement for reliable quantitative analysis of data. For instance, scanning tunneling spectroscopy (STS) and polarized Raman spectroscopy (pRS) would enable one to obtain simultaneous information from a single- and a two-particle probe. In addition, the tunneling tip may be used to locally enhance the electric field of the incoming laser and thus facilitates the exploitation of near-field effects for light scattering.

Near-field effects were used to study molecules with large Raman cross sections \cite{Hartschuh:2003,Steidtner:2008,ZhangZL:2016}. In solids the enhancement effects, specifically the contrast between far and near field are smaller as shown for instance for Si [\onlinecite{Steidtner:2007,Pettinger:2012}] or MoS$_2$ [\onlinecite{Voronine:2017}]. Usually microscope objective lenses having ultra-large working distances are used providing excellent optical images of the surface studied and enough space between sample and lens for the equipment required for near-field enhancement \cite{Hartschuh:2003}. Alternatively, a setup with an on-axis parabola was described recently which due to the very large aperture allows optimal focusing and minimal spot size \cite{Steidtner:2007}. In both cases radially polarized light is required which goes along with losses, and the aperture is always partially obscured, at least for nontransparent materials.

In this paper we report the development of an optical setup which is based on an off-axis parabola as opposed to an on-axis parabola or glass optics. The off-axis paraboloid as an objective allows us to reduce the stray light contribution substantially and to access all polarization combinations of incoming and scattered photons necessary to derive the symmetry components of the Raman response. A tip for atomic force and tunneling microscopy was integrated and demonstrated to work in either mode. This forms the basis for tip-enhanced Raman scattering (TERS) as the next development step.

\section{Description of the setup}

\begin{figure}[tbp]
  \centering
  \includegraphics[width=1.0\columnwidth]{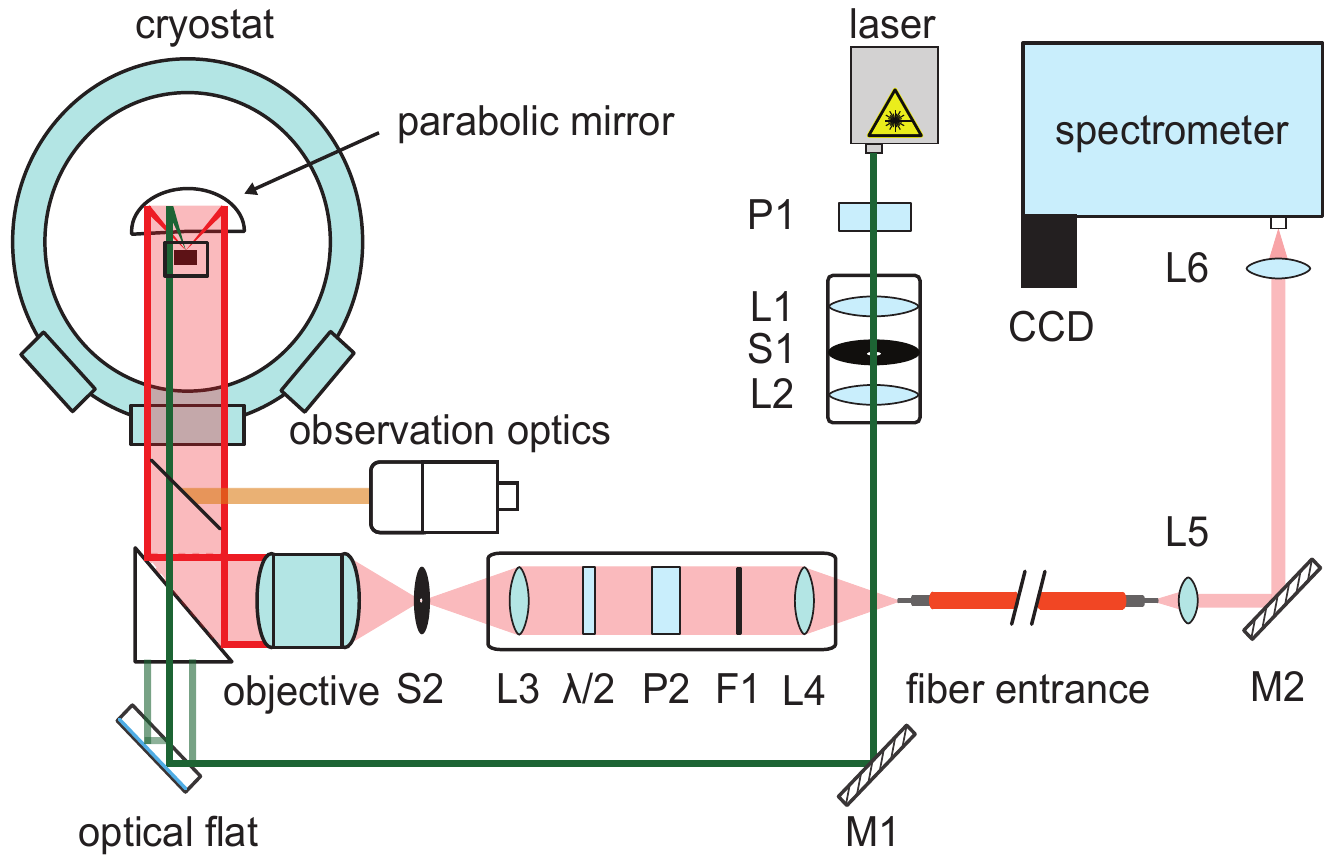}
  \caption{Schematic depiction of the set-up. Sample and parabola are in the center of the cryostat. The polarized (P1) and spatially filtered (L1 S1 L2) incident light passes through a semitransparent mirror into the cryostat, hits the parabola and is focused on the sample (see Fig.~\ref{fig:paraboloid}). The scattered light is reflected into the camera objective lens via the parabola and the semitransparent mirror. The recollected light is spatially filtered (S2). Before entering the fiber different polarization states can be selected ($\lambda/2$, P2) and the elastically scattered light can be filtered out by an edge filter (F1). The fiber transports the light to the spectrometer.
  }
  \label{fig:setup}
\end{figure}

\begin{figure}[tbp]
  \centering
  \includegraphics[width=0.95\columnwidth]{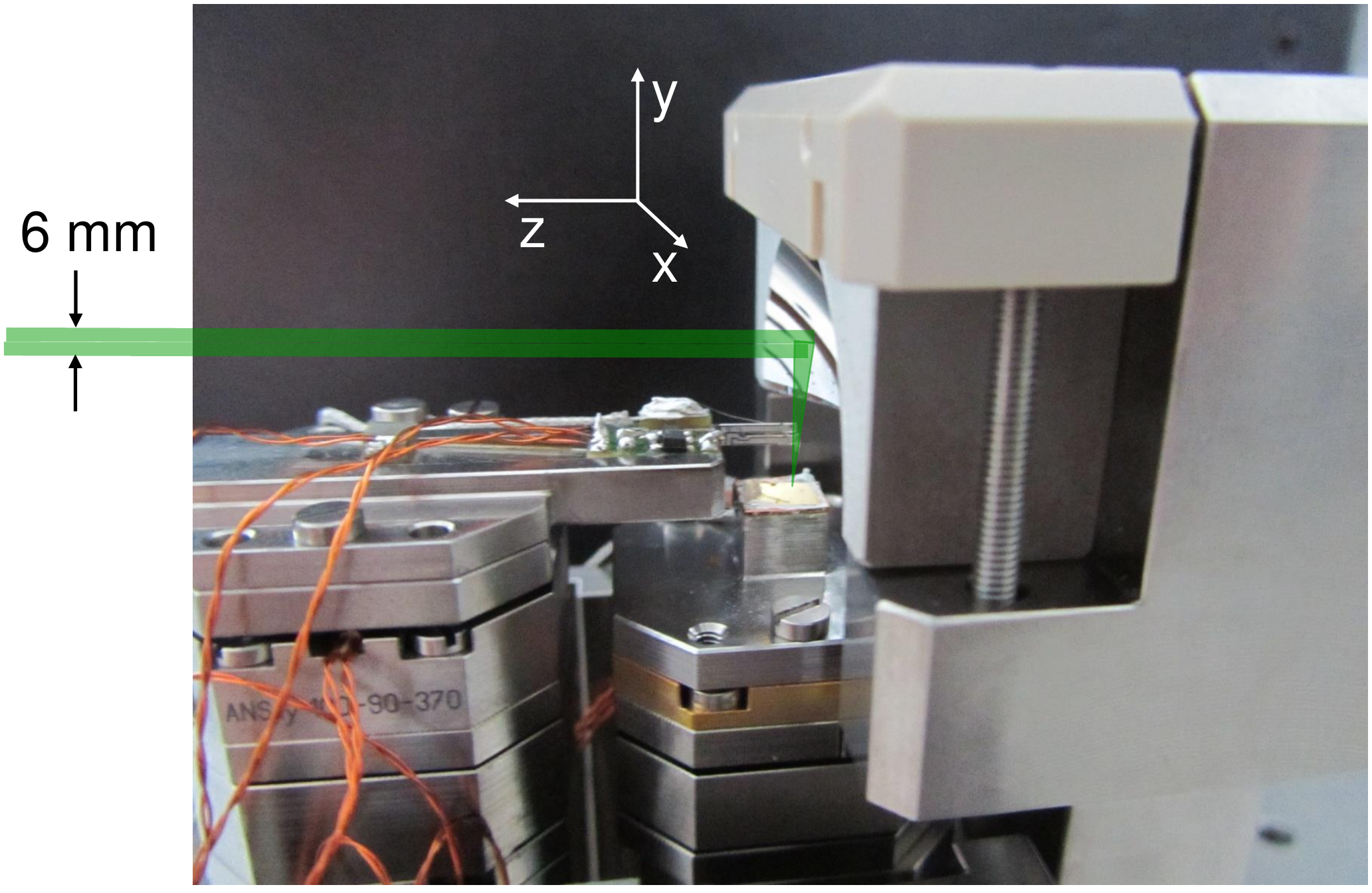}
  \caption{Paraboloid and schematic of the incident laser beam. The beam is focused on the sample surface. The tuning fork for the AFM is visible below the beam. The coordinate system serves for description. The polarizations in the parallel parts of the laser beam and the scattered light are in the $x-y$ plane. The optical axis, the axis of the paraboloid and the $z$-direction coincide. The sample surface and the tip are perpendicular and parallel to the $y$ axis, respectively.
  }
  \label{fig:paraboloid}
\end{figure}

Fig.~\ref{fig:setup} shows our experimental setup. We use diode-pumped solid state (DPSS) lasers emitting either at 532 (Coherent, sapphire SF 532) or 660\,nm (Laser Quantum, ignis 660). The beam is spatially filtered and expanded to a diameter of approximately 6\,mm and has a divergence of $2\times 10^{-4}$\,mrad. The polarized parallel light is directed to one quadrant of the paraboloid ($f=8$\,mm, $D=30$\,mm) and focused on the sample. The sample surface is parallel to the axis of the paraboloid as shown in Fig.~\ref{fig:paraboloid}. This orientation of the surface simplifies the application of the scanning techniques enormously since there is enough space above the surface and the actuators are out of the aperture. In case of optimal alignment the focus on the sample is less than $2\,\mu$m. The scattered light is collected by the paraboloid and ideally parallel since it originates from a point-like source in the focus of the parabola. The parallel light is recollected by a  commercial camera objective lens (Canon, $f=85$\,mm, $f/D=1.4$) and coupled into a multimode fiber (core diameter $d=100\,\mu$m). Finally the scattered light is analyzed by a triple spectrometer (T\,64000 Horiba Jobin Yvon) and recorded with a liquid-nitrogen cooled charge-coupled device (CCD).

Between the camera objective lens and the fiber the light may be filtered spatially for confocal microscopy, and different polarization states may be selected. The desired linear photon polarizations are selected by setting the polarizer at a fixed position and rotate only the $\lambda/2$ achromatic waveplate (see Fig.~\ref{fig:setup}). Since the waveplate has no wedge the focus remains centered on the fiber core and the intensity is polarization independent. The elastic stray light is rejected by an edge filter (Iridian Spectral Technologies). This way the luminescence in the fiber can be reduced to a minimum. The fiber mixes all polarization states, and unpolarized light reaches the spectrometer.

\begin{figure}[tbp]
  \centering
  \includegraphics[width=0.8\columnwidth]{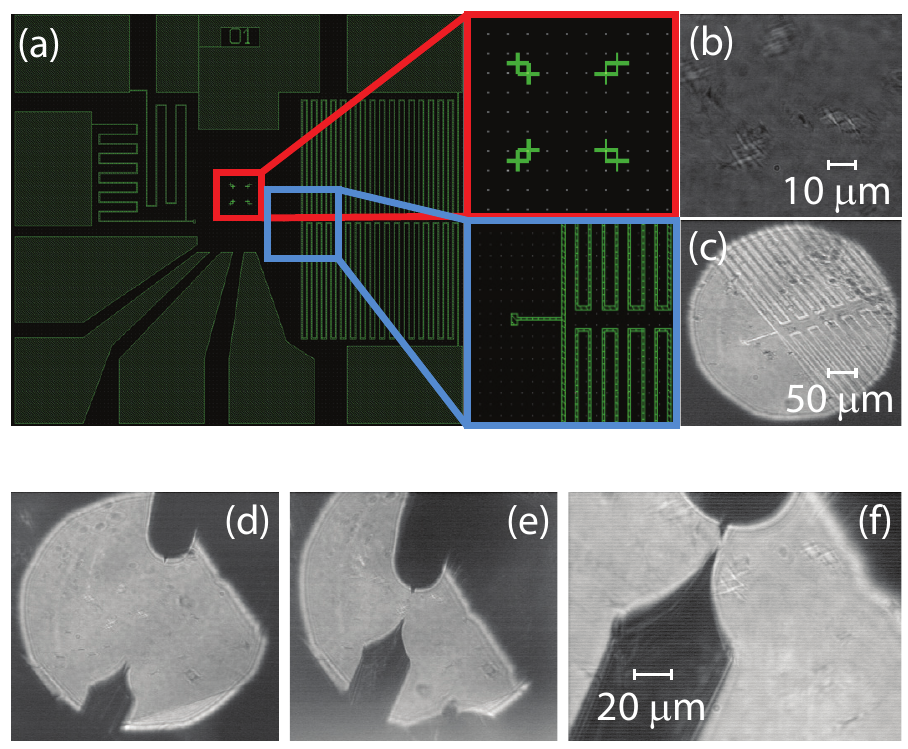}
  \caption{Test of the observation optics. (a) For estimating the resolution we used circuits prepared in the institute (Here Au is deposited on Si for good contrast). (b) and (c) Images obtained via the observation optics. The dimensions are indicated. (d)-(f) The manual approach to a safe distance between the tip and the sample can also be controlled using the observation optics.
  }
  \label{fig:test}
\end{figure}

Use of a parabola yields three major challenges: (i) The exciting light must be accurately aligned parallel to the optical axis of the parabola. (ii) The observation of the sample around the focus of the paraboloid needs to have enough resolution. (iii) The typically weak Raman signal has to be coupled into the fiber.

For aligning the laser parallel to the parabola axis we adopted the method proposed by Lee \cite{Lee:1992} for determining the focus of off-axis parabolic mirrors. For this approach one uses two exactly parallel laser beams which intersect in the focus of the paraboloid only if they are parallel to the optical axis (o.A.). The two parallel beams are generated from one laser beam with a double-sided optical flat (B. Halle Nachfl.). Both sides have an accuracy of $\lambda/20$ and are parallel to each other to within the same accuracy. The resulting error in parallelism is approximately $0.5\,\mu$rad. The flat is made of fused silica and coated with silver on one side. For $\lambda_0>500$\,nm the resulting intensities of the first three reflected beams are 0.040, 0.876, and 0.033. The first and the third beam having approximately the same intensity are used for alignment while the second one is blocked. For the Raman measurements the first and the third beams are blocked.

\begin{figure}[tbp]
  \centering
  \includegraphics[width=0.95\columnwidth]{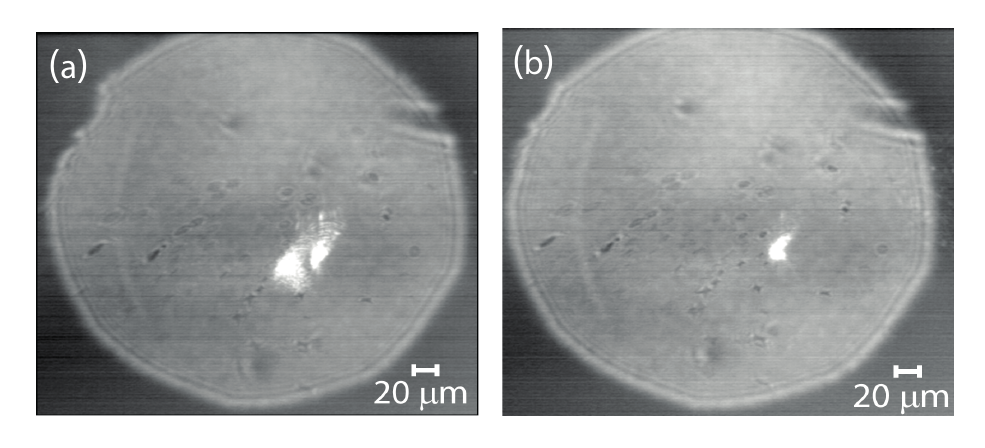}
  \caption{Snapshots of the  alignment procedure. (a) Still two spots are visible meaning either that the beams are not parallel to the parabola axis or that the sample surface is not in the focus. (b) By iteratively adjusting the sample position and the orientation of the laser beams coincidence of the two spots can be achieved.
  }
  \label{fig:focus}
\end{figure}

For the alignment procedure of laser and parabola a high-resolution image of the region around the focus of the parabola and a large field of viewing is needed. We use two independent methods for obtaining those images. The direct observation via a telescope in macro mode (TV Zoom 7000 Navitar, $f=18\dots 108$\,mm) and a camera (not shown in Fig.~\ref{fig:setup}) has a resolution of 10-20\,$\mu$m. The high-resolution observation system (Fig.~\ref{fig:setup}), which uses the parabola as a microscope objective lens and a telesphoto lens (Sigma Optics, $f=70\dots 300$\,mm, $f/D=4.0\dots 5.6$) equipped with another camera, has a resolution of approximately 2\,$\mu$m as can be seen in Fig.~\ref{fig:test}\,(b) and (c). The telephoto lens looks through an aperture of approximately 10\,mm into the center of the parabola via a removable beam splitter. Since the curvature of the parabola across the aperture changes by almost 50\% thus introducing strong astigmatism the images are distorted [Fig.~\ref{fig:test}\,(b) and (c)] but prove sufficient for the intended purpose. Whenever the two beams are not parallel to the optical axis of the paraboloid or the sample surface is out of focus there will be two spots as shown in Fig.~\ref{fig:focus}.

For alignment the orientation of the laser beams and the position of the sample are adjusted iteratively with the parabola fixed. For this purpose the optical flat can be tilted over two axes and shifted left and right (see Fig.~\ref{fig:setup}). For vertical adjustment the entire cryostat can be moved up and down  with a precision and reproducibility of approximately $10\,\mu$m. This adjustment is also important for the low-temperature experiments when the cold part inside contracts while the optical axes need to be maintained. Fig.~\ref{fig:focus} shows two snapshots of the alignment. Once the two spots coincide the overall diameter of the laser illuminated area is smaller than $20\,\mu$m. The Gaussian diameter of the spot is smaller but the exact shape cannot be determined easily for the aberrations.

In the configuration with the sample surface parallel to the axis of the paraboloid the direct reflex of the laser does neither hit the parabola nor the observation optics. Therefore the intensity of the collected elastic stray light is minimal as intended. However, no direct laser light enters the fiber making the alignment next to impossible. For finally coupling the scattered light into the fiber we move a second Si sample into the focus which is tilted by 30$^\circ$ away from the o.A. towards the parabola. Then the direct reflex hits the surface of the paraboloid and is reflected back parallel to the incident beam. This reflex has enough intensity to be coupled into the fiber. Once the transmitted intensity is maximal we return to the surface parallel to the o.A., put it into the focus, and maximize the intensity of the Si $T_{2g}$ phonon.

\begin{figure}[tbp]
  \centering
  \includegraphics[width=0.95\columnwidth]{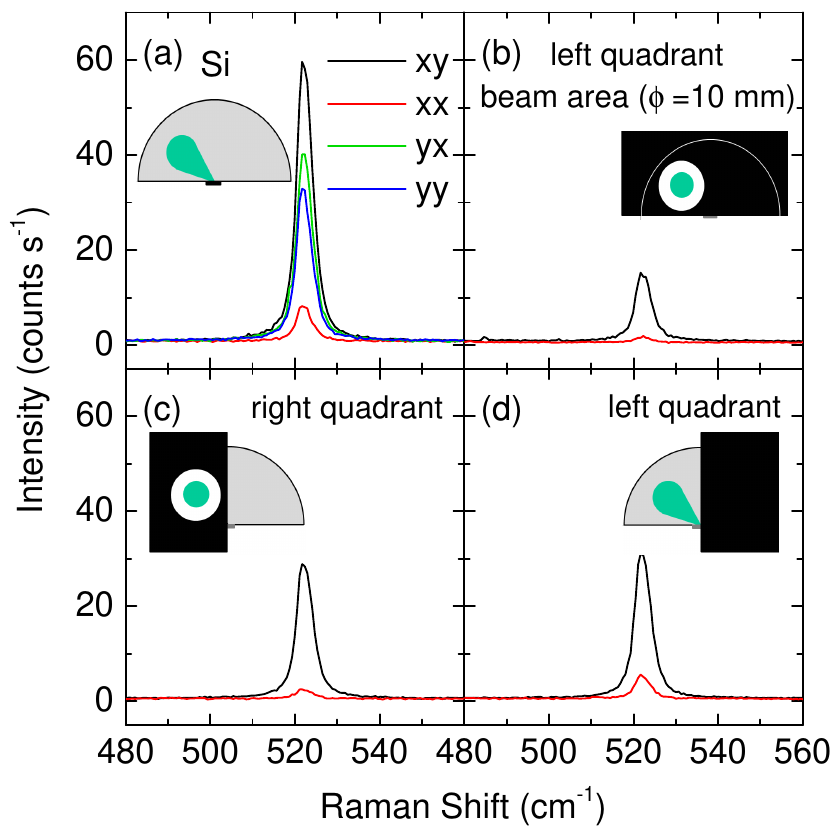}
  \caption{Spectra of Si for different photon polarizations. A \{001\} surface is used here. The polarizations refer to the coordinate system shown in Fig.~\ref{fig:paraboloid}. For $x$- and $y$-polarized photons the projections on the sample surface are close to $[100]$ and $[110]$, respectively, for illumination through the left quadrant as indicated \cite{Hoch:2015}. (a) The scattered light is collected over the full aperture of the parabola. According to the selection rules of Si there is a contrast between $xy$ and $xx$ polarized spectra while the $yy$ and $yx$ spectra have imilar intensity. (b)-(d) The contrast between $xy$ and $xx$ polarized spectra is maintained if parts of the aperture are blocked as indicated.
  }
  \label{fig:Si}
\end{figure}

\section{Experimental results}

With the sample and the fiber aligned we performed polarized Raman scattering measurements so as to experimentally verify the selection rules expected from calculations. One expects that for linearly polarized illumination through one quadrant the linear polarization on the sample surface can be maintained to within a few percent \cite{Hoch:2015}. This result encouraged us to study the effect of various incoming and outgoing polarizations on the selection rules.

The experiments were performed on a \{001\} surface of Si having a strong phonon line at 520\,cm$^{-1}$ with $T_{2g}$ symmetry. Maximal intensity is expected for incoming and outgoing photons, ($\hat{\bf e}_I$;$\hat{\bf e}_S$), polarized along either ($[100]$;$[010]$) or ($[110]$;$[110]$) or equivalent directions. In our experiment the sample's $[100]$ and $[0\bar{1}0]$ axes are oriented parallel to the $x$ and $z$ axes, respectively, of the coordinate system shown in Fig.~\ref{fig:paraboloid}. If only the central part of the parabola is considered photons having $x$ and $y$ polarizations outside the cryostat will be polarized parallel to $[100]$ and $[0\bar{1}0]$, respectively, on the sample. If noncentral parts are used, as shown in Fig.~\ref{fig:Si}\,(a), other orientations will also contribute.

As shown in Fig.~\ref{fig:Si}\,(a) the contrast between polarized (parallel incoming and outgoing polarizations outside the cryostat) and depolarized (perpendicular incoming and outgoing polarizations) spectra is pronounced. More specifically, the ratio of the peak intensities between the $xy$ and the $xx$ spectra is approximately 14 whereas that between $yx$ and $yy$ is close to unity. This result looks counterintuitive at first glance but originates in the different apertures for incoming and scattered light. By and large it is in accordance with the selection rules: The incoming $x$-polarized photons have a polarization close to $[100]$ on the sample even if the light comes from the left quadrant [see inset of Fig.~\ref{fig:Si}\,(a)]. For the scattered photons polarized along the $y$ direction various polarizations between $[\bar{1}10]$ and $[110]$ contribute including $[010]$. Hence the intensity is high. Low intensity can be expected for $x$-polarized scattered photons since the polarization in the sample are close to $[100]$ for major parts of the parabola. If $y$-polarized photons enter through the left quadrant the polarization on the sample is close to $[\bar{1}10]$ and considerations similar to those described above apply. Now the intensity is smaller and the contrast is close to unity.

In a second set of experiments [Fig.~\ref{fig:Si}\,(b)-(d)] the contributions from selected parts of the paraboloid were studied. The incident beam always enters in the middle of the left quadrant. In $xy$ configuration the contributions from the quadrants are approximately equal whereas in $xx$ configuration the intensity collected from the left quadrant is stronger by a factor of 2.5 than that from the right quadrant [Fig.~\ref{fig:Si}\,(c) and (d)]. The intensity ratios for the right and the left quadrant are 14 and 6.2, respectively. If the cross section of the paraboloid is reduced to an aperture of 10\,mm diameter around the laser beam [left quadrant, Fig.~\ref{fig:Si}\,(b)] the intensity is further reduced but the contrast is restored. The reduction of the intensity corresponds approximately to the ratio of the areas of the half paraboloid and the aperture of 10\,mm indicating that the scattered light is in fact collected by a major part of the paraboloid.

\section{Conclusions}

We reported the development of an optical system for low-temperature polarized Raman spectroscopy. The central innovation is an off-axis parabola as an alternative imaging system having a numerical aperture of better than ${\rm N.A.}=0.8$. As its inherent advantage the parabola provides much better mechanical access to the focal point. Therefore additional probes such as AFM and STS may be added. The metallic tips may also be used for near-field spectroscopy. Because of the specific configuration (see Fig.~\ref{fig:paraboloid}) the obscuration is minimal. In contrast to glass objective lenses the parabola may be cooled down to arbitrarily small temperatures. On the other hand, it may be baked and is thus compatible with ultrahigh vacuum (UHV) conditions.

We also installed two different types of observations optics and a K\"ohler-type illumination system (not shown in Fig.~\ref{fig:setup}) allowing us to obtain images of the sample surface. The best resolution and the largest field of view are 2\,$\mu$m and, respectively, $6\times6\,{\rm mm}^2$. The observation optics was used to align the laser beam with respect to the parabola and to control the approach of AFM/STS tips. 

The imaging system on the basis of the off-axis paraboloid enabled us to perform polarized Raman spectroscopy. The polarization leakage can be suppressed to below 7\% upon using the full aperture of the parabola. Therefore  the symmetry components of the scattered light may be derived while a high collection efficiency can be maintained.

\begin{acknowledgments}
We acknowledge useful discussions with C. Pfleiderer. Financial support for the work came from the DFG via the Transregional Collaborative Research Center TRR\,80 and from the European Commission
through the ITN Project FINELUMEN (PITN-GA-2008-215399).
\end{acknowledgments}

\bibliography{D:/!papers/!bib/literatureR2}

\begin{thebibliography}{8}%
\makeatletter
\providecommand \@ifxundefined [1]{%
 \@ifx{#1\undefined}
}%
\providecommand \@ifnum [1]{%
 \ifnum #1\expandafter \@firstoftwo
 \else \expandafter \@secondoftwo
 \fi
}%
\providecommand \@ifx [1]{%
 \ifx #1\expandafter \@firstoftwo
 \else \expandafter \@secondoftwo
 \fi
}%
\providecommand \natexlab [1]{#1}%
\providecommand \enquote  [1]{``#1''}%
\providecommand \bibnamefont  [1]{#1}%
\providecommand \bibfnamefont [1]{#1}%
\providecommand \citenamefont [1]{#1}%
\providecommand \href@noop [0]{\@secondoftwo}%
\providecommand \href [0]{\begingroup \@sanitize@url \@href}%
\providecommand \@href[1]{\@@startlink{#1}\@@href}%
\providecommand \@@href[1]{\endgroup#1\@@endlink}%
\providecommand \@sanitize@url [0]{\catcode `\\12\catcode `\$12\catcode
  `\&12\catcode `\#12\catcode `\^12\catcode `\_12\catcode `\%12\relax}%
\providecommand \@@startlink[1]{}%
\providecommand \@@endlink[0]{}%
\providecommand \url  [0]{\begingroup\@sanitize@url \@url }%
\providecommand \@url [1]{\endgroup\@href {#1}{\urlprefix }}%
\providecommand \urlprefix  [0]{URL }%
\providecommand \Eprint [0]{\href }%
\providecommand \doibase [0]{http://dx.doi.org/}%
\providecommand \selectlanguage [0]{\@gobble}%
\providecommand \bibinfo  [0]{\@secondoftwo}%
\providecommand \bibfield  [0]{\@secondoftwo}%
\providecommand \translation [1]{[#1]}%
\providecommand \BibitemOpen [0]{}%
\providecommand \bibitemStop [0]{}%
\providecommand \bibitemNoStop [0]{.\EOS\space}%
\providecommand \EOS [0]{\spacefactor3000\relax}%
\providecommand \BibitemShut  [1]{\csname bibitem#1\endcsname}%
\let\auto@bib@innerbib\@empty
\bibitem [{\citenamefont {Hartschuh}\ \emph {et~al.}(2003)\citenamefont
  {Hartschuh}, \citenamefont {S\'anchez}, \citenamefont {Xie},\ and\
  \citenamefont {Novotny}}]{Hartschuh:2003}%
  \BibitemOpen
  \bibfield  {author} {\bibinfo {author} {\bibfnamefont {A.}~\bibnamefont
  {Hartschuh}}, \bibinfo {author} {\bibfnamefont {E.~J.}\ \bibnamefont
  {S\'anchez}}, \bibinfo {author} {\bibfnamefont {X.~S.}\ \bibnamefont {Xie}},
  \ and\ \bibinfo {author} {\bibfnamefont {L.}~\bibnamefont {Novotny}},\ }\href
  {\doibase 10.1103/PhysRevLett.90.095503} {\bibfield  {journal} {\bibinfo
  {journal} {Phys. Rev. Lett.}\ }\textbf {\bibinfo {volume} {90}},\ \bibinfo
  {pages} {095503} (\bibinfo {year} {2003})}\BibitemShut {NoStop}%
\bibitem [{\citenamefont {Steidtner}\ and\ \citenamefont
  {Pettinger}(2008)}]{Steidtner:2008}%
  \BibitemOpen
  \bibfield  {author} {\bibinfo {author} {\bibfnamefont {J.}~\bibnamefont
  {Steidtner}}\ and\ \bibinfo {author} {\bibfnamefont {B.}~\bibnamefont
  {Pettinger}},\ }\href {\doibase 10.1103/PhysRevLett.100.236101} {\bibfield
  {journal} {\bibinfo  {journal} {Phys. Rev. Lett.}\ }\textbf {\bibinfo
  {volume} {100}},\ \bibinfo {pages} {236101} (\bibinfo {year}
  {2008})}\BibitemShut {NoStop}%
\bibitem [{\citenamefont {Zhang}\ \emph {et~al.}(2016)\citenamefont {Zhang},
  \citenamefont {Sheng}, \citenamefont {Wang},\ and\ \citenamefont
  {Sun}}]{ZhangZL:2016}%
  \BibitemOpen
  \bibfield  {author} {\bibinfo {author} {\bibfnamefont {Z.}~\bibnamefont
  {Zhang}}, \bibinfo {author} {\bibfnamefont {S.}~\bibnamefont {Sheng}},
  \bibinfo {author} {\bibfnamefont {R.}~\bibnamefont {Wang}}, \ and\ \bibinfo
  {author} {\bibfnamefont {M.}~\bibnamefont {Sun}},\ }\href {\doibase
  10.1021/acs.analchem.6b02093} {\bibfield  {journal} {\bibinfo  {journal}
  {Anal. Chem.}\ }\textbf {\bibinfo {volume} {88}},\ \bibinfo {pages} {9328}
  (\bibinfo {year} {2016})}\BibitemShut {NoStop}%
\bibitem [{\citenamefont {Steidtner}\ and\ \citenamefont
  {Pettinger}(2007)}]{Steidtner:2007}%
  \BibitemOpen
  \bibfield  {author} {\bibinfo {author} {\bibfnamefont {J.}~\bibnamefont
  {Steidtner}}\ and\ \bibinfo {author} {\bibfnamefont {B.}~\bibnamefont
  {Pettinger}},\ }\href {\doibase 10.1063/1.2794227} {\bibfield  {journal}
  {\bibinfo  {journal} {Rev. Sci. Instrum.}\ }\textbf {\bibinfo {volume}
  {78}},\ \bibinfo {pages} {103104} (\bibinfo {year} {2007})}\BibitemShut
  {NoStop}%
\bibitem [{\citenamefont {Pettinger}\ \emph {et~al.}(2012)\citenamefont
  {Pettinger}, \citenamefont {Schambach}, \citenamefont {Villag\'omez},\ and\
  \citenamefont {Scott}}]{Pettinger:2012}%
  \BibitemOpen
  \bibfield  {author} {\bibinfo {author} {\bibfnamefont {B.}~\bibnamefont
  {Pettinger}}, \bibinfo {author} {\bibfnamefont {P.}~\bibnamefont
  {Schambach}}, \bibinfo {author} {\bibfnamefont {C.~J.}\ \bibnamefont
  {Villag\'omez}}, \ and\ \bibinfo {author} {\bibfnamefont {N.}~\bibnamefont
  {Scott}},\ }\href {\doibase 10.1146/annurev-physchem-032511-143807}
  {\bibfield  {journal} {\bibinfo  {journal} {Ann. Rev. Phys. Chem.}\ }\textbf
  {\bibinfo {volume} {63}},\ \bibinfo {pages} {379} (\bibinfo {year}
  {2012})}\BibitemShut {NoStop}%
\bibitem [{\citenamefont {Voronine}\ \emph {et~al.}(2017)\citenamefont
  {Voronine}, \citenamefont {Lu}, \citenamefont {Zhu},\ and\ \citenamefont
  {Krayev}}]{Voronine:2017}%
  \BibitemOpen
  \bibfield  {author} {\bibinfo {author} {\bibfnamefont {D.~V.}\ \bibnamefont
  {Voronine}}, \bibinfo {author} {\bibfnamefont {G.}~\bibnamefont {Lu}},
  \bibinfo {author} {\bibfnamefont {D.}~\bibnamefont {Zhu}}, \ and\ \bibinfo
  {author} {\bibfnamefont {A.}~\bibnamefont {Krayev}},\ }\href {\doibase
  10.1109/JSTQE.2016.2584784} {\bibfield  {journal} {\bibinfo  {journal} {IEEE
  Journal of Selected Topics in Quantum Electronics}\ }\textbf {\bibinfo
  {volume} {23}},\ \bibinfo {pages} {1} (\bibinfo {year} {2017})}\BibitemShut
  {NoStop}%
\bibitem [{\citenamefont {Lee}(1992)}]{Lee:1992}%
  \BibitemOpen
  \bibfield  {author} {\bibinfo {author} {\bibfnamefont {Y.~H.}\ \bibnamefont
  {Lee}},\ }\href {\doibase 10.1117/12.59951} {\bibfield  {journal} {\bibinfo
  {journal} {Opt. Eng.}\ }\textbf {\bibinfo {volume} {31}},\ \bibinfo {pages}
  {2287} (\bibinfo {year} {1992})}\BibitemShut {NoStop}%
\bibitem [{\citenamefont {Hoch}()}]{Hoch:2015}%
  \BibitemOpen
  \bibfield  {author} {\bibinfo {author} {\bibfnamefont {D.}~\bibnamefont
  {Hoch}},\ }\emph {\bibinfo {title} {{Tip-Enhanced Raman Spectroscopy}}},\
  \href@noop {} {Master's thesis}\BibitemShut {NoStop}%
\end{thebibliography}%

\end{document}